\documentclass[%
 reprint,
 amsmath,amssymb,
 aps,prl
]{revtex4-2}

\usepackage{graphicx}
\usepackage{dcolumn}
\usepackage{bm}
\usepackage{color}
\usepackage{hyperref}
\usepackage{nameref}
\usepackage[mathlines]{lineno}
\begin{document}

\preprint{APS/123-QED}




\title{Power Spectra of Velocity Fluctuations in Granular Heap Flow}

\author{Shuchang Yu$^{1}$}
\author{Jin Shang$^{1}$}
\author{Yangrui Chen$^{1}$}
\author{Ran Li$^{2}$}
\author{Quan Chen$^{2}$}

\author{Hui Yang$^{3,2}$}
\email{yangh\_23@sumhs.edu.cn}
\author{Hu Zheng$^{4,5}$}
\email{zhenghu@tongji.edu.cn}
\author{Jie Zhang$^{1,6}$}
\email{jiezhang2012@sjtu.edu.cn }
\address{$^1$ School of Physics and Astronomy, Shanghai Jiao Tong University, Shanghai 200240, China}
\address{$^2$ School of Optical-Electrical and Computer Engineering, University of Shanghai for Science and Technology, Shanghai 200093, China}
\address{$^3$ College of Medical Instrument, Shanghai University of Medicine \& Health Sciences, Shanghai 201318, China}
\address{$^4$ Department of Geotechnical Engineering, College of Civil Engineering, Tongji University, Shanghai, 200092,China}
\address{$^5$ Key Laboratory of Geotechnical and Underground Engineering of Ministry of Education, Tongji University, Shanghai,200092,China}
\address{$^6$ Institute of Natural Sciences, Shanghai Jiao Tong University, Shanghai 200240, China}






\begin{abstract}
This study used Speckle Visibility Spectroscopy to examine velocity fluctuations in a three-dimensional granular heap flow, where the mean velocity profile consists of a fast-flow surface layer and a creep layer beneath. The velocity spectra follow power-law scalings, \(E(f) \propto f^{\alpha}\), with \(\alpha \approx -0.85\)  in the surface flow layer—matching the Self-Organized Criticality (SOC) model with open boundaries ($\alpha \approx -0.95$). In the creep layer, \(\alpha\)  decreases with depth, reaching \(\alpha \approx -1.5\)  at \(\sim 55\)  mean particle diameters, consistent with the SOC model with closed boundaries ($\alpha \approx -1.58$). Analysis of the fluctuation velocity distributions offers additional insights into the microscopic origins of the spectrum’s characteristics. These findings help resolve the long-standing puzzle of flow localization in gravity-driven granular flows despite a constant shear stress-to-pressure ratio throughout the material.

\end{abstract}

\maketitle


\color{blue}\textit{Introduction}
\color{black} --The study of velocity fluctuations has played a crucial role in the development of various scientific fields, including statistical physics~\cite{landau2013statistical}, turbulence\cite{Frisch}, nonlinear dynamic models\cite{1987Self, 1996Self}, and mathematical models of active matter\cite{vicsek1995,toner1995}. 
Granular matter, as a quintessential nonequilibrium system, exhibits a rich spectrum of dynamical behaviors across multiple temporal and spatial scales. Extensive experimental and theoretical investigations under diverse driving conditions have revealed the prevalence of power-law distributions \cite{2002Turbulent,2015Experimental,2012An,2013Eddy,2016Anomalous,2009Dynamical,2008Temporal,2006Jamming,2005Slow,Brian1996Stress,1996Friction, 2016Enstrophy,1991Self,2001Stick,1996Stick,2010Fluctuations,1995Density, 1994Density,2012Relation,1987Self,1996Self}. 
Power-law behavior, for instance, has been observed in turbulent-like motion within shear-driven granular layers \cite{2002Turbulent,2015Experimental,2012An,2013Eddy,2016Anomalous,2022Turbulent,2016Enstrophy}. This phenomenon is intrinsically tied to complex force fluctuations in dense granular media \cite{2009Dynamical,2008Temporal,2006Jamming,2005Slow,Brian1996Stress}, stochastic friction dynamics \cite{1996Friction,1991Self,2001Stick,1996Stick}, density fluctuations in dense granular flows \cite{1995Density,1994Density}, and the nonlinear dynamics underlying self-organized criticality (SOC) in sandpile models \cite{2012Relation,1987Self,1996Self}. Despite these insights, experimentally quantifying the spectral properties of velocity fluctuations and their connection to macroscopic system states remains a significant challenge.

In this study, we employed Speckle Visibility Spectroscopy (SVS) \cite{2005Speckle}, a high-temporal-resolution, non-invasive optical technique widely used in granular systems \cite{2008Avalanche,2000From,2009Particle,2008Statistics,2003Speckle,2010Jamming}. Our analysis of velocity fluctuations in a three-dimensional (3D) steady heap flow revealed two distinct regimes of power spectra: one corresponding to the fast-flowing surface layer and the other to the underlying creep layer.
By connecting the spectral characteristics of these two regimes to the predictions of the self-organized criticality (SOC) model, we aim to offer valuable insights into the longstanding puzzle of surface flow localization in gravity-driven heap flow. This occurs despite a constant shear stress-to-pressure ratio throughout the material.

\begin{figure}[h]
\centering
  \includegraphics[width=8.4cm]{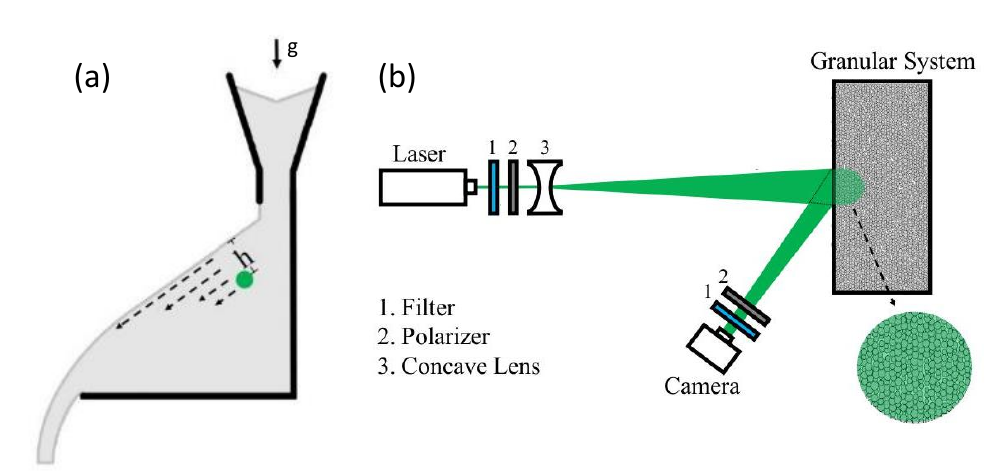}
  \caption{(a) The geometry of a granular heap, including the coordinate system and relevant parameters. (b) A schematic diagram of the Speckle Visibility Spectroscopy setup.}
  \label{fig1}
\end{figure}

\color{blue}\textit{Experimental System}
\color{black} --In this experiment, we investigated granular flows down a heap in a Hele-Shaw-type cell. The cell comprised two parallel, transparent, anti-static acrylic plates (40 cm$\times$25 cm) with a 6 cm gap. The bottom and one side were sealed. Glass beads (0.3$\sim$0.6 mm, $\rho \approx 2.5 g/cm^3$) were continuously poured through a hopper near the sealed side at a rate of $130 g/s$, maintaining steady heap flow with an angle of repose $\theta \approx 26^\circ$. The hopper’s opening matched the gap width for uniform feeding. This setup closely resembles the classic heap-flow model \cite{2004On,2005Speckle}.
Initially empty, the cell gradually formed a heap as particles accumulated. A stable granular flow developed at the surface, with excess particles exiting at the lower corner. All measurements were taken under steady-state conditions. The schematic diagram of the setup is shown in Fig.~\ref{fig1}(a).

\begin{figure}[h]
\centering
  \includegraphics[width=8.4cm]{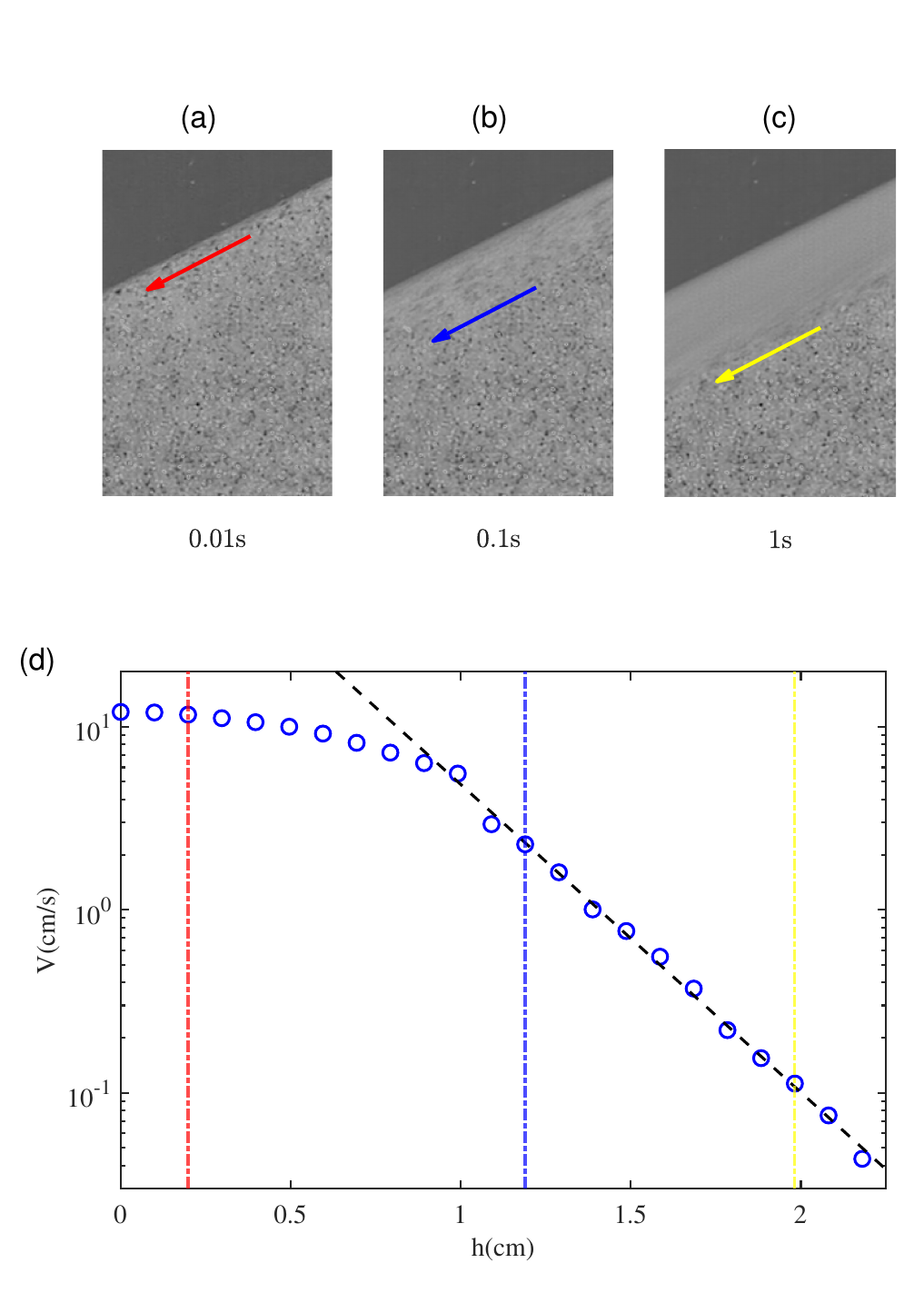}
  \caption{(a-c) Snapshots of a granular heap in a steady flow state. The particles in these images are glass beads with diameters ranging from 0.3 to 0.6 mm. All photos were taken under identical conditions, with the only variation being the shutter speeds. (d) Mean velocity $V$ is plotted against the depth $h$ relative to the surface of the heap.}
  \label{fig2}
\end{figure}

Velocity fluctuations were measured using SVS\cite{2005Speckle}, which determines particle fluctuation velocity through multiple light-scattering events with excellent time resolution.
As shown in Fig~\ref{fig1}(b), a 532 nm, 1 W laser was directed perpendicularly at the acrylic plate. A 532 nm filter and a polarizer ensured consistent polarization and monochromatic illumination. The laser beam, expanded by a concave lens, formed a $\sim0.7$ cm circular spot, penetrating the granular layer.
A $1 \times 4096$, 20 kHz line-array camera, positioned 10 cm from the spot, captured scattered light. The camera lens had a 532 nm filter and polarizer to reduce noise. Before each experiment, the camera's front polarizer was adjusted to minimize light intensity, suppressing reflections from the acrylic plate.

The mean velocity profile in the heap flow varies with depth, with the fastest flow near the surface. We used particle image velocimetry (PIV) to measure this profile, capturing images at a spatial resolution of $960 \times 1280$.
Since velocity changes by orders of magnitude with depth, we adjusted frame rates (1000--0.1 Hz) to match variations. A small fraction of dyed particles enhanced image contrast for velocity measurements.

Figure~\ref{fig2}(a) shows snapshots of a stable granular flow with particles fed from the right. The images confirm that the mean flow has reached a steady state, as indicated by the unchanged position of the flow surface. Additionally, the progressive boundary profiles separating the creeping and seemingly stationary layers remain parallel to the flow surface. This aligns with the well-known characteristics of gravity-driven surface flow, where the shear stress-to-pressure ratio remains constant throughout the material\cite{schall2010shear}.
Colored arrows in Fig.~\ref{fig2}(a-c) indicate velocity direction and measurement positions. Figure~\ref{fig2}(d) plots average particle velocity parallel to the surface against depth. The x-axis represents distance from the surface, while the y-axis shows velocity on a logarithmic scale. The black dashed line fits an exponential function, while red, blue, and yellow dashed lines correspond to sampling frequencies of 100 Hz, 10 Hz, and 1 Hz, matching Fig.~\ref{fig2}(a-c).
Consistent with earlier findings \cite{J2012Experimental}, the mean velocity profile reveals two distinct regions: particles in the surface layer exhibit faster flow, while those in the lower layer display creeping motion, with the transition depth occurring at approximately 1 cm.


\color{blue}\textit{Velocity Power Spectra}
\color{black} --To better understand the temporal fluctuations in granular flow, we quantify velocity variations by computing the corresponding power spectrum:
$E(f) = \left| \int_{}^{} \delta v(t) e^{-i 2 \pi f t} \, dt \right|^2.$
The spectrum analysis shows the strength of velocity fluctuations at various time scales and depths of the heap flow.

\begin{figure}[h]
\centering
  \includegraphics[width=8.4cm]{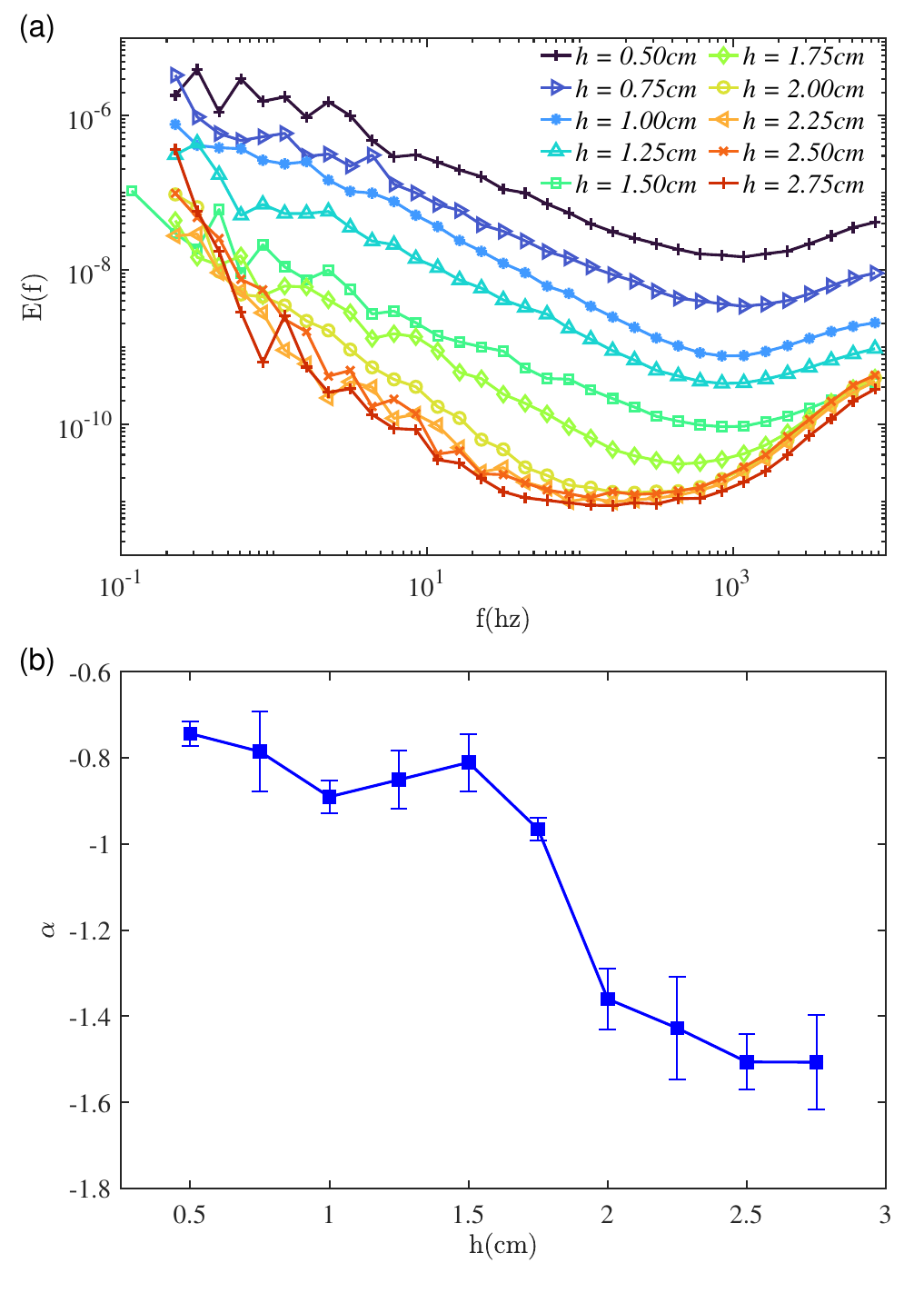}
  \caption{(a) The power spectra of the fluctuation velocity display a power-law behavior, which can be fitted as \( f^{\alpha} \). The fitting exponent \( \alpha \) varies with depth, as shown by the colors and legend. (b) The exponent \( \alpha \) is plotted as a function of depth \( h \).
}
  \label{fig3}
\end{figure}

Figure~\ref{fig3}(a) shows the velocity power spectra of the heap flow at various depths, revealing clear power-law characteristics over a broad frequency range. These power-law behaviors suggest that the system's dynamics exhibit self-similar characteristics in the temporal domain. The slope of the power spectra in the log-log plot varies with depth. To quantify the corresponding power-law exponents, we performed power-law fitting on the spectral curves using
\(
E(f) \propto f^{\alpha}
\) 
and extracted the exponent $\alpha$ for each curve.
The increase in spectrum curves at high frequencies, such as above approximately \(10^3\) Hz, is significantly influenced by the static structure factor at the grain scale due to the discrete characteristics of granular materials\cite{2022Turbulent}.

Figure~\ref{fig3}(b) shows the exponent $\alpha$ as a function of depth $h$, where $h$ represents the vertical distance from the heap flow surface. The curve exhibits two distinct regimes: For depths between $h = 0.5$ cm and $1.5$ cm, the exponent remains approximately constant at $\alpha \approx -0.85$. Beyond $h = 1.5$ cm, extending to $2.75$ cm, $\alpha$ gradually decreases with depth, reaching a constant value of $\alpha \approx -1.5$ around a depth of $h=2.5$ cm. This two-regime characteristic resembles the mean velocity profile, though with a slightly larger transition depth of $1.5$ cm. This discrepancy may result from the finite size of the light illumination spot in the SVS technique introduced earlier. To further interpret the results in Fig.~\ref{fig3}(b), we propose a possible explanation based on an SOC model\cite{1987Self, 1996Self}.

An important physical quantity introduced within the theory of SOC \cite{1987Self, 1996Self} is the energy dissipation rate, defined as the total number of sliding events occurring throughout the system per unit time. This quantity, denoted as \(F(t)\), is presumably a function of time \(t\). It has been reported that in a 2D sandpile of open boundaries, the spectrum of \(F(t)\) follows a power-law distribution, with a power-law exponent of approximately \(-0.95\). Let us assume \(\mathcal{F}\)  represents the Fourier transform, hence,
\(
\left| \mathcal{F}\{ F(t) \} \right|^2 \propto f^{-0.95}
\). 
To better understand the measured power spectra in the shallow layer, we assume that the velocity \( V_{\text{soc}}(\tau) \) is directly proportional to the total number of sliding events that occur within a coarse-grained time interval \(\tau\). Moreover, we assume that the sliding dynamics are consistent with the SOC model\cite{1987Self, 1996Self},
\(
V\text(\tau) \sim \langle F(t)\rangle_{\tau},
\)
where the notation $\langle \rangle_{\tau}$ refers to the coarse-graining over the time interval $\tau$.

 Coarse-graining over $\tau$ does not influence the spectral properties of $F(t)$ at frequencies below approximately $1/\tau$. Consequently, the spectrum of the velocity \( V(\tau) \) is determined by
\(
\left| \mathcal{F}\{ V(\tau))\} \right|^2\sim \left| \mathcal{F}\{ \langle F(t)\rangle_{\tau} \} \right|^2 \propto f^{-0.95}.
\)
The numerical result of the power-law exponent, \(-0.95\), closely aligns with the experimentally measured exponent of \(\alpha\approx-0.85\). 
This agreement suggests that the localized surface flow layer is governed by the process of SOC with open boundaries \cite{1987Self, 1996Self}.

It can be observed from Fig.~\ref{fig3}(b) that when the depth exceeds 1.5 cm (around 33 mean particle diameters), significant changes occur in the fluctuation characteristics of the granular flow. A deviation of approximately 0.5 cm was observed between the transition depths identified by the spectral exponent and those determined by the velocity profile. Considering the centimeter-level spatial resolution of the SVS technology used, which has a surface laser spot diameter of about 0.7 cm, it is hypothesized that the measurement scattering effect may have increased the detection area. This could lead to the mixing of signals from the creeping zone with those from the surface flow components, resulting in a change in depth determination.

It is intriguing to note that in the SOC model\cite{1987Self, 1996Self}, the exact value of $\alpha$ depends on the boundary conditions: for closed boundaries, the 2D SOC numerical model yields $\alpha\approx -1.58$ , which closely matches the constant value of $\alpha\approx -1.5$ at a depth of 2.5 cm (approximately 55 mean particle diameters) and beyond, as shown in Fig.~\ref{fig3}(b). 
Furthermore, the transition from open to closed boundaries aligns well with the experimental setup illustrated in Fig.~\ref{fig1}(a). In the context of the SOC model, the gradual adjustment of the value of $\alpha$ corresponds to a smooth transition from open boundaries to closed ones.


To our knowledge, only two previous granular experiments, both in 2D, have measured power spectra of nonaffine displacement in wavevector space \cite{2012An, 2022Turbulent}. The first, involving disordered binary disks under cyclic shear \cite{2022Turbulent}, analyzes displacement fields from successive cycles. While interesting, this approach differs significantly from ours. The second experiment, which examines quasi-statically sheared binary disks, reports a power-law exponent of $\alpha \approx-5/3$ \cite{2012An}, resembling Kolmogorov turbulence \cite{Frisch}.
Based on insights from Fig.~\ref{fig3}, this result may be reinterpreted through the SOC model with closed boundaries\cite{1987Self, 1996Self}. However, caution is needed when converting spectra from wavevector to frequency space due to the quasistatic nature of shear \cite{2012An}. A similar issue arises in previous numerical studies \cite{2002Turbulent, Oyama-PRL-2019, 2016Anomalous, 2016Enstrophy}, where quasi-static shear prevents true dynamics. Moreover, reported values of $\alpha$ vary across these studies despite their similar systems. Given these complications and the periodic boundary conditions \cite{2002Turbulent, Oyama-PRL-2019, 2016Anomalous, 2016Enstrophy}, they fall outside the scope of our present study.

\begin{figure}[h]
\centering
  \includegraphics[width=8.4cm]{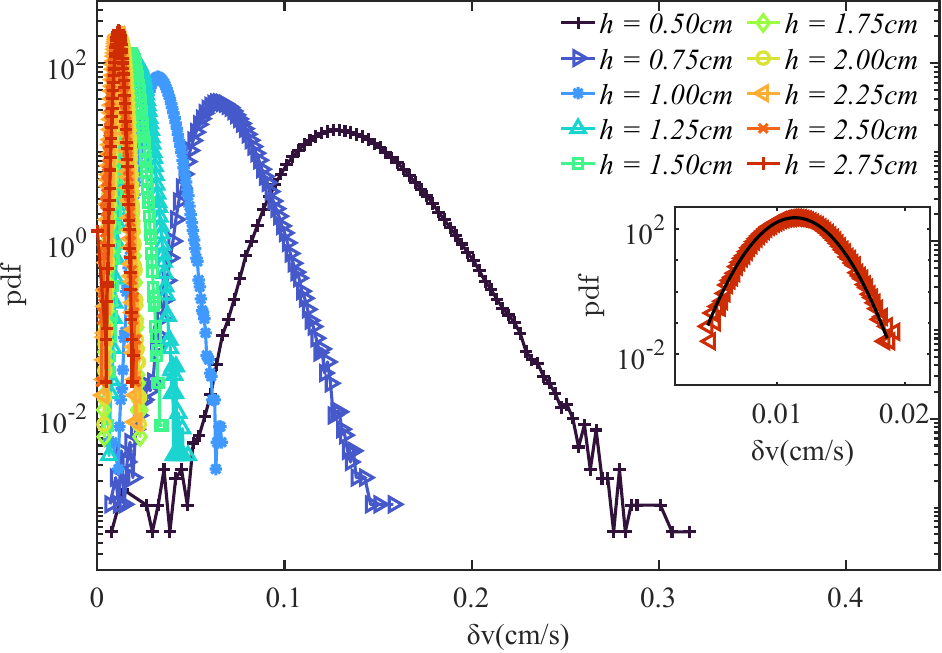}
  \caption{The probability distribution functions (pdf) of fluctuation velocity at various depths. The inset highlights the velocity distribution at a depth of 2.75 cm, with a black solid line representing a Gaussian fit.
}
  \label{fig4nu}
\end{figure}

\color{blue}\textit{Connection with Surface Flow Localization}
\color{black} -- The results in Fig.~\ref{fig3} provide valuable insights into the localized shear flow in the surface layer of the heap flow, as shown in Fig.~\ref{fig2}. Flow localization in heap flow is a representative case of many similar inhomogeneous flow profiles observed in geological and engineering contexts. Our findings suggest that the same underlying physics governs both surface and creep flows, albeit under different boundary conditions.

For a heap with grains continuously added at the upper end and exiting at the lower end, the boundary conditions closely resemble the open boundaries in the SOC model \cite{1987Self, 1996Self}. In contrast, at sufficiently deep layers, creep flows naturally obey the closed boundary conditions of the SOC model \cite{1987Self, 1996Self}. The boundary conditions in the intermediate layers can be interpolated between these two limits.

This explanation also extends to homogeneous flow profiles observed in flows on inclined planes, where the system aligns well with the open-boundary conditions in the SOC model \cite{1987Self, 1996Self}. This results in nearly uniform velocity profiles under various conditions, as investigated in the classical work of MiDi \cite{2004On}.

Flow localization and the related phenomenon of shear localization are ubiquitous in granular matter \cite{schall2010shear}, soft glassy materials \cite{Fielding}, complex fluids \cite{Divoux}, and metallic glasses \cite{Greer}. Whether a universal physical mechanism governs these phenomena remains an open question. Our findings highlight the critical role of boundary conditions -- an aspect that has received little attention in the past but may be worth exploring in future investigations.

\color{blue}\textit{Fluctuation Velocity Distributions}
\color{black} --Studies on granular systems \cite{2013Signatures,2022Fluctuation,2004The,2006Stress,John2008Velocity,2004Diffusion} often show significant deviations from Gaussian velocity fluctuations in macroscopic flows due to inelastic collisions, frictional dissipation, and complex interparticle interactions.
As shown in Fig.~\ref{fig4nu}, velocity fluctuations near the surface are more skewed and gradually approach a Gaussian form when depth increases. The inset highlights the velocity distribution at a depth of 2.75 cm, with a black solid line representing a Gaussian fit.

In the surface flow layer, gravity-driven acceleration dominates due to the open boundary at the lower end, converting potential energy into directional kinetic energy. Frequent inelastic collisions and enduring contacts dissipate some energy, but dissipation lags behind energy input, causing continued acceleration and relatively significant non-Gaussian distributions. Due to the gradual change of the open to a closed boundary condition at greater depths, an increase in particle density restricts movement, transitioning the flow into a creep regime dominated by enduring contacts and friction. The velocity fluctuation spectrum's power-law exponent decreases, and the distribution trends toward Gaussian, indicating the onset of random diffusion. Notice that the changes in the statistics of velocity fluctuations in Fig.~\ref{fig4nu} are continuous, aligning with the governing physics of SOC\cite{1987Self, 1996Self} as revealed by the spectrum results mentioned above.

\color{blue}\textit{Conclusions}
\color{black} --We use Speckle Visibility Spectroscopy to measure velocity fluctuations at various depths in granular heap flow, which has a fast flow surface layer and a creep flow layer beneath, as revealed by the mean velocity profile from the PIV measurement. 
In the surface layer (0.5–1.5 cm), the energy spectra follow \( E(f) \propto f^{\alpha} \) with \( \alpha \approx -0.85 \); in the creep layer below 1.5 cm, \( \alpha \) gradually decreases with depth, eventually reaching a constant \( \alpha \approx -1.5 \) when the depth is around $h=2.5$ cm. The spectra in both layers are consistent with the predictions of the SOC model, albeit subject to different boundary conditions.
These findings provide new insights into the long-standing mystery of flow localization in gravity-driven heap flow.

\color{blue}{\it Acknowledgments} \color{black} -- SY, JS, YC and JZ acknowledge the support of the NSFC (No.11974238 and No.12274291) and the Shanghai Municipal Education Commission Innovation Program under No. 2021-01-07-00-02-E00138. SY, JS, YC and JZ also acknowledge the support from the SJTU Student Innovation Center. HZ acknowledges the support of the NSFC (No.42277156).

\nocite{*}

\bibliography{apssamp}

\end{document}